\documentclass[fleqn,11pt]{article}

\usepackage{amsfonts,amsmath,amssymb}
\usepackage{latexsym,color,cite}

\usepackage[utf8]{inputenc}
\usepackage{graphicx}
\usepackage{amsfonts,amssymb,cite}
\usepackage{graphicx}

\def\noi{\noindent}

\newcommand{\Title}[1]{\noi {{\Large\bf #1}}\\[1ex]}

\def\Aunames#1{\noi{\bf #1}}
\def\au#1{${}^{#1}$}
\def\Addresses#1{\medskip\noi \protect
	\begin{description}\itemsep -3pt {\it #1} \end{description}}
\def\adr#1#2{\item[${}^{#1}$]{\it #2}}

\newcommand{\Abstract}[1]{\vskip 2mm \begin{center}
        \parbox{16.4cm}{\small\noi #1} \end{center}\medskip}
\newcommand{\foom}[1]{\protect\footnotemark[#1]}
\newcommand{\foox}[2]{\footnotetext[#1]{#2}\addtocounter{footnote}{1}}
\def\email#1#2{\footnotetext[#1]{e-mail: #2}\addtocounter{footnote}{1}}

\def\nq{\hspace*{-1em}}
\def\nqq{\hspace*{-2em}}
\def\nhq{\hspace*{-0.5em}}

\def\qq{\qquad}
\def\cm{\hspace*{1cm}}
\def\inch{\hspace*{1in}}

\usepackage{color}
\def\red#1{{\color{red} #1}}
\def\blue#1{{\color{blue} #1}}
\def\Funding#1{\subsection*{Funding} #1}

\def\ConflictThey{\subsection*{Conflict of interest} 
	The authors declare that they have no conflicts of interest.}

\def\Jl#1#2{#1 {\bf #2},\ }

\def\ApJ#1 {\Jl{Astroph. J.}{#1}}
\def\CQG#1 {\Jl{Class. Quantum Grav.}{#1}}
\def\DAN#1 {\Jl{Dokl. AN SSSR}{#1}}
\def\GC#1 {\Jl{Grav. Cosmol.}{#1}}
\def\GRG#1 {\Jl{Gen. Rel. Grav.}{#1}}
\def\IJMPD#1 {\Jl{Int. J. Mod. Phys. D}{#1}}
\def\JETF#1 {\Jl{Zh. Eksp. Teor. Fiz.}{#1}}
\def\JETP#1 {\Jl{Sov. Phys. JETP}{#1}}
\def\JHEP#1 {\Jl{JHEP}{#1}}
\def\JMP#1 {\Jl{J. Math. Phys.}{#1}}
\def\NPB#1 {\Jl{Nucl. Phys. B}{#1}}
\def\NP#1 {\Jl{Nucl. Phys.}{#1}}
\def\PLA#1 {\Jl{Phys. Lett. A}{#1}}
\def\PLB#1 {\Jl{Phys. Lett. B}{#1}}
\def\PRD#1 {\Jl{Phys. Rev. D}{#1}}
\def\PRL#1 {\Jl{Phys. Rev. Lett.}{#1}}

\def\lal{&&\nqq {}}
\def\eq{Eq.\,}

\def\beq{\begin{equation}}
\def\eeq{\end{equation}}
\def\besub{\begin{subequations}}
\def\esub{\end{subequations}}
\def\bear{\begin{eqnarray}}
\def\bearr{\begin{eqnarray} \lal}
\def\ear{\end{eqnarray}}
\def\earn{\nonumber \end{eqnarray}}

\def\nnn{\nonumber\\ \lal }

\def\yyy{\\[5pt] \lal }

\def\dst{\displaystyle}
\def\tst{\textstyle}
\def\fracd#1#2{{\dst\frac{#1}{#2}}}
\def\fract#1#2{{\tst\frac{#1}{#2}}}
\def\Half{{\fracd{1}{2}}}
\def\half{{\fract{1}{2}}}

\def\e{{\,\rm e}}

\def\arg{\mathop{\rm arg}\nolimits}

\def\diag{\mathop{\rm diag}\nolimits}

\def\const{{\rm const}}

\def\ep{\epsilon}

\def\then{\ \Rightarrow\ }

\begin{document}
\twocolumn[
\thispagestyle{empty}

\Title{Magnetic mirror stars in five dimensions\foom 1}

\Aunames{Kirill A. Bronnikov,\au{a,b,c,2} Sergei V. Bolokhov,\au{b,3} and Milena V. Skvortsova,\au{b,4}}
	
\Addresses{\small
\adr a	{Center of Gravitation and Fundamental Metrology, ROSTEST, 
		Ozyornaya ul. 46, Moscow, 119361, Russia}
\adr b	{Institute of Gravitation and Cosmology, RUDN University, 
		ul. Miklukho-Maklaya 6, Moscow, 117198, Russia}
\adr c 	{National Research Nuclear University ``MEPhI'', 
		Kashirskoe sh. 31, Moscow, 115409, Russia}
		}

%%\Dates{February 10, 2025}{February 10, 2025}{February 20, 2025}
\medskip

\Abstract
   {We discuss a class of solutions of multidimensional gravity which are formally
    related to black-hole solutions but can observationally look like compact stars whose
    surface reflects back all particles or signals getting there. Some particular examples
    of such solutions are presented and studied, including those with a magnetic field 
    in Maxwell or nonlinear electrodynamics (NED) in five dimensions. 
    For NED as a possible source for magnetic mirror stars, we formulate a methodology 
    of solving the 5D Einstein-NED equations and point out the conditions under which 
    there always exist mirror star solutions. We also note that some of the Einstein-Maxwell
    solutions under consideration are discussed in the literature and called 
    ``topological stars'' due to the circular topology of the fifth dimension. }

] % ===========================================   
\foox 1 {Extended version of a talk presented at the 
	18th Russian Gravitational Conference (RUSGRAV-18), Kazan, 25-29 November 2024.}
\email 2 {kb20@yandex.ru}  
\email 3 {boloh@rambler.ru}  
\email 4 {milenas577@mail.ru}  

{  %%   au-defs
% -------------------------
\def\red#1{{\color{red} #1}}
\def\blue#1{{\color{blue} #1}}
\def\eqn#1{\eq\eqref{#1}}
\def\rf{\eqref}
% -------------------------
\def\mn{_{\mu\nu}}
\def\MN{^{\mu\nu}}
\def\mN{_\mu^\nu}
\def\nM{_\nu^\mu}
% ---------------------
\def\M{{\mathbb M}}
\def\R{{\mathbb R}}
\def\Z{{\mathbb Z}}
\def\cF{{\mathcal F}}
\def\cL{{\mathcal L}}
\def\cO{{\mathcal O}}
\def\og{{\overline g}}
\def\oq{{\overline q}}
\def\oR{{\overline R}}
% --------------------------
\def\dxi{\delta\xi}
\def\Veff{V_{\rm eff}}
\def\oalpha{\overline{\alpha}}
\def\obeta{\overline{\beta}}
\def\ogamma{\overline{\gamma}}
% -------------------------
\def\chg{\ \leftrightarrow\ }

\def\GR{general relativity}
\def\sph{spherically symmetric}
\def\ssph{static, spherically symmetric}
\def\bh{black hole}
\def\bhs{black holes}
\def\wh{wormhole}
\def\whs{wormholes}
\def\asflat{asymptotically flat}
\def\emag{electromagnetic}
\def\mult{multidimensional}
\def\pb{perturbation}
\def\pbs{perturbations}
\def\Scw{Schwarz\-schild}
\def\Schr{Schr\"odinger}
\def\RN{Reiss\-ner-Nord\-str\"om}

% ============================= sec 1
\section{Introduction}
% =============================

   Multidimensional gravity is well known to supply a great variety of predictions of possible 
   compact objects which may in principle exist in Nature. In addition to a vast collection of 
   \bh\ models (see for reviews, e.g., [1--5], these are wormholes 
   [6--9], boson stars, gravastars, etc. which possess many features 
   of interest but are actually direct extensions or generalizations of their 4D counterparts.
   
   Meanwhile, one can also discuss such hypothetic objects whose very existence is only possible 
   due to extra dimensions. The corresponding solutions to multidimensional gravity equations 
   may be obtained from \bh\ solutions where one should mutually replace time and one of 
   the extra  dimensions  \cite{bb-kb95a, bb-kb95b}. Such a replacement becomes possible because 
   the equations ``do not know'' which of the coordinates is time and which is considered to 
   be ``extra.'' Such solutions are almost equally numerous with \bh\ ones: a necessary condition 
   for their emergence is the existence of a suitable 1D subspace in the multidimensional 
   space-time. And, as argued in  \cite{bb-kb95a, bb-kb95b}, if the suitable extra dimension is compact,
   spacelike, and sufficiently small, then the surface that has been an event horizon in the \bh\ 
   counterpart of our object, now acquires the properties of a perfectly reflecting surface.
   Therefore, we suggest to call such objects {\sl mirror stars.}
   
   The possible occurrence of reflection phenomena related to compact objects is rather actively
   discussed in the astrophysical context. Various kinds of echoes are predicted in \bh\ and \wh\
   configurations [12--15]. Thus, explicit restrictions have been recently 
   obtained \cite{bb-25-lim}  on totally reflective exotic compact objects whose physical surfaces 
   are located at a radius $r = r_h (1 + \ep)$ from a would-be event horizon radius $r_h$ and 
   $\ep < 10^{-3}$.
   
   In the present paper, we discuss some examples of mirror star solutions in the framework 
   of 5D gravity. The electro-(magneto-)vacuum solutions presented here are special cases 
   of \ssph\ solutions obtained and discussed previously in \cite{bb-kb95a, bb-kb95b} but deserve a 
   more detailed study. In addition, we present two mirror-star solutions sourced by nonlinear
   electrodynamics. We have shown that there is a class of magneto-vacuum solutions that are 
   stable under \sph\ \pbs \cite{bb-we25}, and this stability range corresponds to the above-mentioned
   ``compactness parameter'' $\ep \in (0, 0.42)$ (where the upper bound is obtained numerically). 
   Let us note that these objects that we suggest to call  ``magnetic mirror stars'' were recently
   discussed in a number of papers under the name of ``topological stars'' [18--20]
   (as justified by the circular topology of the 5th dimension). In particular, in \cite{bb-tops3} such 
   objects were shown to be stable under nonspherical \pbs, which adds interest in their 
   further exploration. We also noticed that the the stability range under \sph\ \pbs\ obtained in
   our study and in \cite{bb-tops2, bb-tops3} are slightly different, and we hope to discuss this 
   issue in our forthcoming paper \cite{bb-we25}.
   
   This paper is organized as follows. In Section 2 we begin with a discussion of the simplest 
   example of a mirror-star solution, namely, a 5D version of the \Scw\ metric with 
   $g_{tt} = \const$. It illustrates the main feature of such space-times, the reflection property 
   of their inner boundary that replaces the \Scw\ horizon. In Section 3 we consider 5D
   electro/magneto-vacuum solutions with a Maxwell field as a source, and in Section 4 we
   present two examples of mirror-star solutions sourced by nonlinear electrodynamics. 
   Section 5 is a conclusion. 
      
% ============================= sec 2
\section{5D \Scw\ space-time as a mirror star}
% =============================

   A direct product of the 4D \Scw\ manifold and a flat extra manifold with any signature 
   is known to possess a zero \mult\ Ricci tensor and, hence satisfies the \mult\ Einstein 
   equations. In such an extra manifold, one may single out a 1D subspace and ascribe the 
   \Scw\ factor $1- 2m/r$ to the corresponding extra coordinate, leaving ``our'' time 
   coordinate with the factor 1. After this substitution the whole manifold will be still
   Ricci-flat since this property does not depend on our interpretation of different 
   submanifolds. So, let us consider, as the simplest example, such a 5D Ricci-flat manifold
   with the metric, which, following \cite{bb-kb95a, bb-kb95b}, may be called {\it T-Schwarzschild}, 
\bearr	  \label{bb-TScw}
	 ds_5^2 = dt^2  - \bigg(1- \frac{2m}{r}\bigg)^{-1} dr^2 - r^2 d\Omega^2
\nnn \inch	 
	 		+ \bigg(1- \frac{2m}{r}\bigg)\eta_v dv^2. 
\ear
  having a {\it T-horizon\/} at $r=2m$, where $\eta_v = \pm 1$ corresponds to a 
  timelike ($+1$) or spacelike ($-1$) nature of the fifth coordinate $v$. 

  At a transition from $r>2m$ to $r < 2m$, the signs of $g_{rr}$ and $g_{vv}$ change 
  simultaneously, but the whole situation strongly depends on $\eta_v$. Thus, if $\eta_v =1$, 
  so that the $v$ coordinate is timelike at large $r$, the overall signature of the 5D metric 
  remains the same at small $r$, while at $\eta_v =-1$ the $v$ coordinate is spacelike at 
  large $r$ and becomes timelike at small $r$, simultaneously with $r$ itself. Thus at large 
  $r$ we have one timelike and four spacelike coordinates, while at small $r$ there are 
  ``three times'' and only two spacelike directions. Recall that everything happens in a 
  Ricci-flat space-time, and, as is easily verified, there is no curvature singularity at $r = 2m$.

  In the case $\eta_v=1$, the surface $r = 2m$ is similar to a ``usual'' \Scw\ horizon, but in the
  $(r,v)$ 2D subspace, and the metric can be extended from $r> 2m$ to $r < 2m$, for 
  example, using Kruskal-like coordinates, resulting in the well-known Kruskal diagram. 
  However, if the extra dimension $v$ should be invisible by 4D observers, we must assume 
  by the Kaluza-Klein recipe that it is compact and very small, and this is achieved by
  identifying some values of $v$, say, $v=0$ and $v= 2\pi\ell$ with $\ell$ the compactification
  length. This operation singles out some narrow wedge-like sector in the Kruskal diagram
  ar $r > 2m$ and a similar sector at $r < 2m$, whose only common point coincides with 
  the horizon intersection point. We conclude that, with a compact $v$ axis, the surface 
  $r = 2m$ becomes a naked conical singularity.
  
  Let us look what happens in the case $\eta_v=-1$. In an attempt to jointly study the regions 
  $r > 2m$ and $r < 2m$, let us introduce such coordinates that the metric is obvioustly 
  nonsingular at $r =2m$. For the metric (\ref{bb-TScw}), we can introduce such coordinates 
  close to $r = 2m$ as follows:
\bearr                                    \label{bb-Thor}
		r - 2 m = \frac{x^2+y^2}{8m}, \qq   v = 4m \arctan \Big(y/x\Big), 
\nnn
	`ds^2_2 (r,v) \approx   \frac{r - 2m}{2m} dv^2 + \frac{2m}{r-2m}dr^2 
\nnn \inch	
	= dx^2 + dy^2.
\ear
  We see that the metric of the $(r,v)$ surface is locally flat at the point $r=2m$, which itself 
  has become the origin $x=y=0$, and the $v$ coordinate behaves there as a polar angle.

  A similar transformation has been applied to some static cylindrically 
  symmetric Einstein-Maxwell solutions in \cite{bb-br79}, see also \cite{bb-cy20}. It is closely
  related to a conformal mapping using the analytic function $\log z,\ \ z=x+ {\rm i}y$ 
  in the complex plane, and our $v$ coordinate is then proportional to $\arg\,z$. 
 
  We conclude that the $(r,v)$ surface near $r = 2m$ behaves in general like the 
  Riemann surface for the function $\log z$ that consists of a finite or infinite 
  (if the range of $v$ is infinite) number of sheets, and $r=2m$ is their branch point.   
  To make it regular, we assume $0\leq v < 2\pi \ell$, identifying $v=0$ and $v=2\pi \ell$,
  where $\ell$ is a compactification length at large $r$. As a result, the location $r = 2m$ 
  becomes a center of symmetry in the regular $(r,v)$ surface. Thus the $(r,v)$ surface  
  has the shape of a chemical test tube that has a constant thickness at large $r$
  and a regular end at $r = 2m$. It is thus clear that a transition across $r = 2m$,
  to a region with another signature of the metric is impossible. 

  If we consider a radial geodesic in 5D space, along which $v = \const$, then its
  projection onto the ($r,v$) surface reaches the point $r=2m$ and, due to its regularity, 
  passes it and returns to larger $r$ but now with a changed value of $v$, say, 
  $v = \pi\ell$ instead of $v=0$. A point particle moving along this geodesic will thus 
  leave a particular 4D section of space-time. However, assuming the extreme smallness of $l$ 
  it is more reasonable to consider a quantum particle whose wave function
  is $v$-independent (since a dependence on $v$ would create its huge mass). Then
  this particle should remain visible to a 4D observer who will conclude that it 
  has been reflected from the surface $r = 2m$. In other words, the T-horizon $r=2m$
  is observed as a mirror surface and can be called a {\it mirror horizon}. 
  
  It is natural to expect a similar behavior of all T-horizons appearing in \mult\ space-times
  with a spacelike extra dimension. 
  
% ====================================================
\section{Black holes and mirror stars among 5D Einstein-Maxwell\\ space-times} 
% =====================================================

  In this section we will briefly discuss some special \ssph\ solutions of 5D \GR\ with a 
  Maxwell field (the Lagrangian is $L = - F = - F_{AB} F^{AB}$ as the only source of gravity 
  Since these solutions are special cases of those considered in \cite{bb-kb95a, bb-kb95b}
  and even more general classes of \mult\ solutions discussed in \cite{bb-bh5, bb-bbm97}, we 
  do not describe their derivation but only present their expressions and discuss some of 
  their properties.   
  
  We write the \ssph\ 5D metric in the general form 
\bearr           \label{bb-ds5}
		ds_5^2 = g_{AB} dx^A dx^B
\nnn	\ \ \	
		=\e^{2\gamma} dt^2 - \e^{2\alpha} du^2 - \e^{2\beta} d\Omega^2 
				+\eta_v \e^{2\xi} dv^2,
\ear  
  where $u$ is an arbitrary radial coordinate, 
  $d\Omega^2 = d\theta^2 + \sin^2\theta d\varphi^2$, and $v$ is, as before, the fifth
   coordinate, with $\eta_v = \pm 1$. Capital Latin indices refer to all five coordinates, 
   counted as $\{t,u,\theta,\varphi, v\}=\{0,1,2,3,5\}$, while 4D indices are Greek.
  
  In such 5D space-time, besides the radial electric ($F_{01} = -F_{10}$) and 
  magnetic ($F_{23} = -F_{32}$) fields, there can be one more sort of \emag\ field in accord
  with the symmetry of \rf{bb-ds5}, the one whose vector potential has the component $A_5(u)$,
  which leads to nonzero $F_{15} = - F_{51}$. Such an \emag\ component behaves like a
  scalar in 4D space and may be called a ``quasiscalar'' \cite{bb-kb95a}. 
  
  The Maxwell equations for the electric and quasiscalar fields have the solutions
\bearr
		F^{01} = q_e/ \sqrt{|g|}, \qq F^{51} = q_s/ \sqrt{|g|},
\nnn \qq		
		 \sqrt{|g|} = \e^{\alpha + 2\beta + \gamma + \xi} \sin\theta,
\ear  
  where $q_e$ and $q_s$ are integration constants interpreted as the electric and 
  quasiscalar charges, respectively. For a magnetic field we have, as usual, the component 
  $A_3 = q_m \cos\theta$ of the 5-potential, hence $F_{23} = q_m \sin \theta$, where 
  $q_m$ is a magnetic charge. One can jointly write
\bearr                     \label{bb-FF}
		\Big\{F_{01}F^{01}, \ F_{23}F^{23}, \  F_{15}F^{15} \Big\}
\nnn \nhq	
		= \Big\{ -q_e^2 \e^{-4\beta - 2\xi},\
				 q_m^2 \e^{-4\beta},\ 
			- \eta_v q_s^2 \e^{-4\beta - 2\gamma} \Big\}.
\ear  
 
   Under the assumption that only one of the charges is nonzero, general solutions to 
   the Einstein equations with the metric \rf{bb-ds5} have been obtained (see  \cite{bb-kb95a,bb-kb95b} 
   and references therein), and most of them contain naked singularities. Only some of them
   have different kinds of horizons, and it is such special cases that are of more physical
   interest. Let us present them here.
   
% ----------------------------   
   \paragraph{Electric field.} With an electric charge $q = q_e$, we obtain a \bh\ solution with 
   the metric    
\bearr                             \label{bb-bh-e}
		ds_5^2 = \frac{1-2k/x}{(1+p/x)^2}dt^2
\nnn 		
	- \bigg(1+ \frac px\bigg) \bigg[\frac{dx^2}{1-2k/x}  + x^2 d\Omega^2 - \eta_v dv^2 \bigg]
\ear 
   where the integration constants $k>0$ and $p>0$ are related to $q$ and the 
   \Scw\ mass $m$ as 
\beq   		\label{bb-mq-e}
   		m = k+p,\qq     q^2 = \frac 34 p (p+2k).
\eeq   
   The solution is \asflat\ at $x\to\infty$, where $x \approx r = \sqrt{-g_{22}}$, the 
   spherical radius. The surface $x = 2k$ is an event horizon.    
   The electric field strength is given by
\beq
			{\vec E}^2 = F_{01} F^{10} = \frac {q^2}{x^4 (1 + p/x)^3}.
\eeq   
    
% ----------------------------   
  \paragraph{Magnetic field.} In this case ($q = q_m$) there emerge two \asflat\ solutions
  with horizons:  the \bh\ one, with the event horizon at $x = 2k$, such that
\bearr		  \nhq                         \label{bb-bh-m}
		ds_5^2 = \frac{1\!- \!2k/x}{1+p/x} dt^2- \Big(1+ \frac px\Big)^{\!2}\ 
					\!\! \bigg[\frac{dx^2}{1\! - \! 2k/x} + x^2 d\Omega^2\bigg] 
\nnn \inch					
					+ \frac {\eta_v dv^2}{1 + p/x},
\ear
   and a solution with a mirror horizon at $x = 2k$:
\bearr		                           \label{bb-th-m}
		ds_5^2 = \frac{dt^2}{1+p/x} - \bigg(1+ \frac px\bigg)^{\!2}\  
					\bigg[\frac{dx^2}{1-2k/x}  + x^2 d\Omega^2\bigg] 
\nnn \inch							
					+ \frac {1-2k/x}{1 + p/x} \eta_v dv^2.
\ear   
   In both cases, there are two positive parameters $k$ and $p$, and, as in \rf{bb-mq-e}, 
   we have the charge $q^2 = \frac 34 p (p+2k)$, while for the \Scw\ masses we obtain,
   respectively,
\bearr
		{\rm for} \ \rf{bb-bh-m}:\ \ m_{\rm bh} = k+p/2; 
\nnn
		{\rm for} \ \rf{bb-th-m}:\ \ m_{\rm mirr} = p/2.
\ear 
   For the magnetic induction we obtain in both cases, according to \rf{bb-FF},
\beq
		{\vec B}^2 = F_{23} F^{23} = \frac {q^2}{(x + p)^4} = \frac{q^2}{r^4},
\eeq   
    since in these metrics the spherical radius is simply $r = x +p$.    
   
 % -------------------------------------  
\paragraph{Quasiscalar fields.} In this case ($q = q_s$), we obtain two different 
    solutions with T-horizons for $\eta_v = 1$ and $\eta_v = -1$:
\bearr                             \label{bb-ds-s+}
	\eta_v = 1, \ \ \ 
	ds_5^2 = \Big(1 + \frac px\Big) \bigg[ dt^2 - \frac {dx^2}{1- 2k/x} 
\nnn \cm\cm	                
                - x^2 d\Omega^2 \bigg] +  \frac{1-2k/x}{(1+p/x)^2} dv^2,
\yyy                             \label{bb-ds-s-}
	\eta_v = - 1, \ \ \ \ 
	ds_5^2 = \Big(1 + \frac px\Big) \bigg[ dt^2  - \frac {dx^2}{1- 2k/x} 
\nnn \cm\qq
	- x^2 d\Omega^2 \bigg]
                -  \frac{Q^2(1-2k/x)}{(k + p - kp/x)^2} dv^2,
\ear
   As is clear from the previous discussion, only the metric \rf{bb-ds-s-} may be ascribed to 
   a mirror star. Meanwhile, the expression of $g_{tt}$ in these both metrics, since $p > 0$, 
   shows that the \Scw\ mass $m = - p/2$ is negative, and such solutions do not seem to be of 
   much physical interest.    

\paragraph{A magnetic mirror star.}     
   Let us discuss the most relevant of the above solutions, that for a magnetic 
   mirror star with the metric \rf{bb-th-m}, $\eta_v = -1$. It is convenient to rewrite the metric 
   in terms of the spherical radius $r = x + p$:
\bearr               \label{bb-ds-m-r}
		ds_5^2 = \bigg(1 - \frac pr\bigg) dt^2- \frac {dr^2}{(1-p/r)[1 - (p+2k)/r]} 
\nnn \cm\cm
	- r^2 d\Omega^2 -  \bigg(1 - \frac{p+2k}{r}\bigg)dv^2.
\ear   
  This metric contains two parameters, $p>0$ and $k>0$, related to the mass $m$ and the 
  magnetic charge $q =q_m$ by 
 \bearr      \label {k->q}
 		p = 2m = \sqrt{k^2 + 4q^2/3} -k \ 
\nnn \cm 		
 		\then \ k = \frac{q^2 - 3m^2}{3m}.
 \ear 
  Since $k > 0$, the solution requires a sufficiently large magnetic charge: $q^2 > 3m^2$.

  The metric \rf{bb-ds-m-r} coincides with the one considered in [18--20] as 
  describing a ``topological star'' if we change the notations as follows:
\beq
		   r_B = p +2k, 	\qq r_S = p.
\eeq  
  Thus $r_B$ is the mirror horizon radius, and $r_S$ is the \Scw\ radius of a would-be \bh\
  with the same mass. We have always $r_B > r_S$, but they can be arbitrarily close since
  $k$ may be small. Though, the observational constraint obtained in \cite{bb-25-lim} make 
  too close values of these radii ($\ep = r_B/r_S -1 \lesssim 10^{-3}$) highly unlikely. 

  It can be shown \cite{bb-we25} that the present magnetic solution is stable under radial 
  \pbs\ if its parameters satisfy the condition
\beq  		\label{bb-st-cond}
   		\frac pk > s_0 \approx 4.736, 
\eeq   
  where the upper bound has been obtained numerically.
  Rewriting this stability condition in terms of $m$ and $q$, we obtain  a restriction on 
  the ratio $q^2/m^2$. Combined with the inequality $q^2 > 3m^2$, it leads to
  the following range of radially stable mirror stars:
\beq                            \label{bb-stab-rg}
			1 < \frac {q^2}{3 m^2} < 1 + \frac {2}{s_0} \approx 1.42.
\eeq   
  
  A comparison with \cite{bb-tops2, bb-tops3} reveals some disagreement. Thus, in those papers,
  the authors divide the range of the ratio $r_B/r_S$ (which is equal to $q^2/(3m^2)$ in 
  our notations) into two sectors: Type I ($r_B/r_S > 3/2$) and Type II ($r_B/r_S \leq 3/2$),
  depending on the existence of photon spheres. As is mentioned in the introduction 
  of \cite{bb-tops3}, ``the Type I sector was found to be free of instabilities,'' while according 
  to \rf{bb-stab-rg}, the whole Type II sector is unstable.
  
  One more correction is in order.	In the same introduction of \cite{bb-tops3} it is written that 
  ``the metric compactified to 4D diverges at $r = r_B$, which corresponds to a curvature 
  singularity, but the solution is perfectly regular in the five-dimensional uplift.''
  In fact, the 4D subspace (which, by the way, in not compactified) is regular at $r = r_B$,
  where there is a \wh\ throat. If there were no fifth dimension, it would describe a \wh,
  and only the fact that $g_{55}$ vanishes there, makes this would-be throat a nontraversable
  mirror surface. 
    
% ===========================================
\section{Mirror stars from nonlinear\\ electrodynamics (NED)}   
% ===========================================

  Consider, instead of the Maxwell field, a material source of gravity in 5D \GR\ in
  the form of NED with the Lagrangian $L(\cF)$, where $\cF = F_{AB} F^{AB}$,
  It is well known that in 4D the corresponding \ssph\ solutions with electric or magnetic
  charges are quite easily found in terms of the \Scw\ radial coordinate $r$, either for a 
  prescribed form of $L(\cF)$ or by choosing a desired metric function 
  $A(r)= \e^{2\gamma}$ [24--29]; certain difficulties only emerge in 
  attempts to find dyonic solutions with combined electric and magnetic charges (see 
  \cite{bb-N-dyon, bb-N5} and references therein). Unlike that, in 5D the problem looks much 
  more involved. We will try to obtain some solutions of interest, assuming for     
  certainty and for simplicity that there is only a radial magnetic field, so that $F_{AB}$ 
  components other than $F_{23} = - F_{32}$ are zero.
  
  Thus we are considering the 5D Einstein equations with the stress-energy tensor
  corresponding to such a magnetic field:
\beq       \label{bb-SET}  \nq\,
		T^A_B = \Half \diag\Big(L, L, L- 2\cF L_cF, L- 2\cF L_\cF, L\Big),
\eeq   
  where $L_\cF = dL/d\cF$. Thus we have the equality $T^0_0 = T^1_1 = T^5_5$. It is also
  clear that the invariant $\cF$ has the same form as in 4D: 
\beq       \label{bb-cF}
		\cF = 2 F_{23} F^{23} = \frac{q_m^2}{r^4(u)},     
\eeq
  where, as before, $q_m$ is the magnetic charge, and $r = \e^\beta$ is the spherical radius.
  
  Let us present the nonzero components of the Ricci tensor for the metric \rf{bb-ds5} 
  without fixing the radial coordinate $u$ (the prime denotes $d/du$): 
\bearr            \label{bb-Ric-gen}
		R^0_0 = -\e^{-2\alpha}\Big[\gamma'' +\gamma'(2\beta'+\gamma'+\xi'-\alpha')\Big],
\nnn		
  		R^1_1 = - \e^{-2\alpha}\Big[2\beta'' + \gamma'' + \xi'' 
    			+ 2\beta'^2 + \gamma'^2 + \xi'^2
\nnn \inch    			
    			 - \alpha'(2\beta' + \gamma' + \xi')\Big],
\nnn
  		R^2_2 = R^3_3 = \e^{-2\beta}- \e^{-2\alpha}\Big[\beta'' 
\nnn \inch  		
  		+ \beta'(2\beta'+\gamma'+\xi'-\alpha') \Big],
\nnn
		R^5_5 = - \e^{-2\alpha}\Big[\xi'' + \xi'(2\beta'+\gamma'+\xi'-\alpha') \Big],	
\ear  
  Note that putting $\xi = \const$, we obtain the Ricci tensor components $R\mN$ of
  the 4D section of \rf{bb-ds5}.  It is also useful to present the Einstein tensor component 
  $G^1_1 = R^1_1 - \half R$ since the corresponding Einstein equation is first-order and 
  is an integral  of the others which are second-order:
\bearr            \label{bb-G11}
		G^1_1 = - \e^{-2\beta} + \e^{-2\alpha}
				\big(\beta'^2 + 2\beta'\gamma' 
\nnn \inch				
				+ 2\beta'\xi'  + \gamma'\xi'\big).
\ear 

  As follows from \rf{bb-Ric-gen}, the relation $R^0_0 = R^5_5$ (a consequence of 
  \rf{bb-SET}) is easily integrated giving
\beq            \label{bb-int0-5}
		\xi'-\gamma' = N \e^{-(2\beta + \gamma + \xi- \alpha)}, \qq N = \const.
\eeq    
    Similarly, the relation $R^1_1 = R^5_5$ yields
\beq                \label{bb-1-5} 
	2\beta'' + 2\beta'^2 + \gamma'' + \gamma'^2 - (\alpha'+\xi') (2\beta' + \gamma') =0.
\eeq    
    
    So far, we did not fix the choice of the radial coordinate $u$. Now, let us use the
    coordinate $u = x$ defined by the condition $\alpha + \xi =0$ (an analogue of the so-called
    quasiglobal coordinate in 4D defined by $\alpha+\gamma =0$). With this assumption, 
    \eq \rf{bb-1-5} gives  (the prime now denotes $d/dx$)
\beq                \label{bb-1-5q} 
	2\beta'' + 2\beta'^2 + \gamma'' + \gamma'^2 =0,
\eeq        
   an equation giving a more or less transparent way to choose a promising form
   of the unknowns. Let us change the notations as follows:
\beq
		\e^{2\xi} = A(x), \qq \e^\beta = r(x), \qq  \e^\gamma = \Gamma(x),
\eeq     
   and keep in mind that a mirror horizon to be sought for must be a regular zero of $A(x)$. 
   Equations \rf{bb-int0-5} and \rf{bb-1-5q} are rewritten as follows:
\bearr                 \label{bb-0-5q} 
		A' - 2\frac{\Gamma'}{\Gamma} A = \frac {N}{\Gamma r^2},
\yyy                    \label{bb-1-5qq}
		\frac{\Gamma''}{\Gamma} + 2\frac {r''}{r} =0, 		
\ear   
   Equation \rf{bb-0-5q} may be solved as a linear first-order equation with respect to $A(x)$,
   giving 
\beq               \label{bb-A1}
			A(x) = N \Gamma^2 \int \frac{\Gamma^{-3}} {r^2} dx. 
\eeq
   Requiring asymptotic flatness at $x = \infty$, we can suppose $A(\infty) = \Gamma(\infty) =1$,
   and then rewrite \rf{bb-A1} as  
\beq               \label{bb-A2}
			A(x) = \Gamma^2 \bigg(1 - N \int_x^\infty \frac{\Gamma^{-3}}{r^2} dx\bigg),
\eeq    
   where $N > 0$ provides $A(x) < \Gamma^2(x)$, as is required for obtaining a mirror horizon
   at which $A =0$ while $\Gamma > 0$.
   
   More than that, suppose that the functions $\Gamma(x)$ and $r(x)$ are known, \eqn{bb-1-5qq} 
   is satisfied, and $\Gamma(x) > 0$ at $x > x_0$, where $x_0$ is some fixed value of the 
   coordinate $x$.  Then, since the integrand in \eqn{bb-A2} is positive, the constant $N$ can 
   be chosen in such a way that $A(x_0) =0$. In other words, for any particular solution 
   to \eqn{bb-1-5qq} satisfying the asymptotic flatness conditions and any $x_0$ we can choose
   the integration constant $N$ so that $x=x_0$ is a horizon.
   
   We thus obtain the following scheme for finding mirror star solutions: (i) find a solution 
   to \eqn {bb-1-5qq} to determine suitable $r(x)$ and $\Gamma(x)$; (ii) substitute them to 
   \eqn{bb-A2} to determine $A(x)$; (iii) use the now known metric coefficients to find $\cF(x)$
   and $L(x)$ from the Einstein equations; (iv) using \rf{bb-cF}, it is now easy to find the 
   NED Lagrangian $L(\cF)$.  
   
   In what follows we will present two examples of such solutions.
  
\paragraph{Example 1.}  Let us suppose 
\beq
		r(x) = \sqrt{x^2 + a^2} \ \then\  \frac{r''}{r} = \frac{a^2}{(x^2+a^2)^2}. 
\eeq   
   Then from \eq \rf{bb-1-5qq} we can obtain $\Gamma(x)$ as follows:
\bearr
		\Gamma(x) = \Gamma_0 \sqrt{x^2 + a^2}\sin (\xi - \xi_0), 
\nnn		
		\xi(x) = \sqrt{3} \arctan\frac xa, \qq  \Gamma_0, \xi_0 = \const.
\ear		   
  The condition $\Gamma(\infty) = 1$ determines the values of $\xi_0$ and $\Gamma_0$:
\bearr                    \label{bb-1-Ga}
		\xi_0 = \frac{\sqrt{3}\pi}{2}, \qq  \Gamma_0 = - \frac 1{\sqrt{3} a}\ \then \
\nnn		
	 	\Gamma(x) = \frac{\sqrt{x^2 + a^2}}{\sqrt{3} a}\,
  					\sin \Big(\frac{\sqrt{3}\pi}{2} - \sqrt 3 \arctan \frac xa\Big).
\ear 
  At large $x$, we obtain $\Gamma(x) \approx 1 - a^2/(3x^2)$, and the absence of a term 
  proportional to $1/r$ means that this object has a zero \Scw\ mass.
    
  To determine the metric completely, it remains to find $A(x)$ according to \rf{bb-A2},
  while $\e^{2\xi} = 1/A(x)$.  The integral in \rf{bb-A2} can be calculated analytically in terms 
  of $z = \pi/2 - \arctan (x/a)$, but  its expression is rather cumbersome, containing a combination 
  of a few hypergeometric functions, and we do not believe that it makes sense to present
  it here, as well as the expression for $L(\cF)$ obtainable using the expression \rf{bb-G11}
  (according to \rf{bb-SET}, $L(\cF) = - 2G^1_1$). Instead, let us briefly describe the main qualitative
  features of the solution.
  
  To begin with, while $r(x)$ is an even function (as is the case, for example, in symmetric
  \wh\ solutions), the expression for $\Gamma(x)$ is asymmetric, as illustrated in Fig.\,\ref{bb-f-Ga1}.
  At some $x=x_s < 0$ (depending on $a$), we have $\Gamma(x) =0$. On the other hand, the 
  integral in the expression \rf{bb-A2} for $A(x)$ diverges at $x = x_s$ in such a way that 
  $A(x_s)$ is finite, but its sign depends on the value of $N$. There is a (depending on $a$) 
  critical value $N_{\rm cr}$ of $N$, such that at $N \leq N_{\rm cr}$ we have $A(x_s) \geq 0$,
  and the zero value of $g_{00} = \Gamma^2$ makes $x = x_s$ a naked singularity.
  Unlike that, if $N > N_{\rm cr}$, then $A(x)$ reaches zero at some $x = x_h > x_s$,
  which means that such $x = x_h$ is a mirror horizon.  
  
 % --------------------------------------------------- fig 1
\begin{figure}
\centering
\includegraphics[width=7cm]{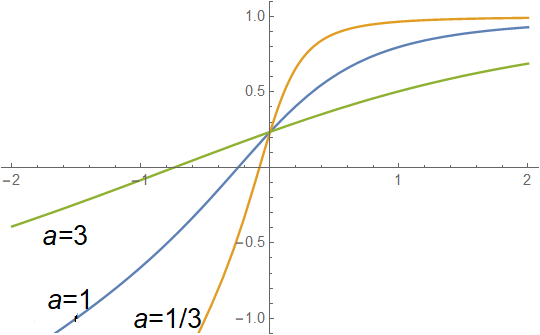}
\caption{\small 
	 The behavior of $\Gamma(x)$ in Example 1 with different values of the parameter $a$ }
\label{bb-f-Ga1}
\end{figure}
% ------------------------------------------------------
 
\paragraph{Example 2.}  Let us suppose 
\beq  			\label{bb-2-Ga}
			\Gamma(x) = 1 - \frac {m}{r(x)},
\eeq
   where $m$ has the meaning of the \Scw\ mass. 
   It then turns out that all quantities are conveniently expressed in terms of $r$. 
   Equation \rf{bb-1-5} leads to an expression of $x$ in terms of $r$:
\beq
		x = r - \frac{m^2}{2(2r-m)} + m \log \frac{2r-m}{r}.
\eeq      
   The resulting \asflat\ metric reads 
\bearr
	   ds_5^2 = \Big(1-\frac mr\Big)^2 dt^2 
       				- \frac{16 r^4 dr^2}{(2r -m)^4 A(r)} 
\nnn \inch       				
       				- r^2 d\Omega^2 - A(r) dv^2, 
\yyy
       A(r) = \Big(1-\dfrac mr\Big)^2 
        		\bigg\{ 1 - 2 N \bigg[\dfrac{4 m^2 - 10 m r + 5 r^2}{(r-m)^2 (2r-m)} 
\nnn \inch        		
        		+ \dfrac 6m \log \dfrac{2 (r-m)}{2r-m} \bigg] \bigg\}. 
\ear        	
    Using \eq \rf{bb-G11} for $G^1_1$, we obtain the following expression for $L(\cF(r))$:
\bearr 
		L(\cF(r)) = \frac 1{8 m r^8}
		\Bigg[4 N m (m - 2 r)^3 
\nnn \qq \times		
		(4 m^2\! - 6 m r\! - 3 r^2) +  m^3 (2 m^4 - 18 m^3 r 
\nnn 		
		+ 63 m^2 r^2 - 104 m r^3 + 72 r^4) 
		  - 12 N (2 r - m )^4 
\nnn \qq \times			  
		  (2 m^2 - 2 m r - r^2) \log \frac{2 (r-m)}{2 r-m}\Bigg] 
\ear       	

   The behavior of the functions $A(r)$ and $L(\cF(r))$ with different values of the integration 
   constant $N$ is illustrated in Fig.\,2. It is evident that zeros of $A(r)$ that correspond to 
   mirror horizons are observed at $r = r_h > m$ and any $N > 0$, and at small $N$ this horizon 
   approaches the surface $r=m$. One can also notice that the Lagrangian function $L(\cF)$
   tends to minus infinity as $r\to m$, but this value is outside the relevant range $r > r_h$.
   
   At large $r$, $L(\cF)$ rapidly decreases by the law $L = \dfrac {3m(3m-N)}{r^4} + O(r^{-5})$, 
   which means that, provided that $N < 3m$, the NED source of this mirror star configuration has 
   a correct Maxwell limit, with a magnetic charge given by $q^2 =  3m(3m-N)/2$. Thus it makes 
   sense to consider our solution only with $N < 3m$.  
       
 % --------------------------------------------------- fig 2
\begin{figure*}
\centering
\includegraphics[width=7cm]{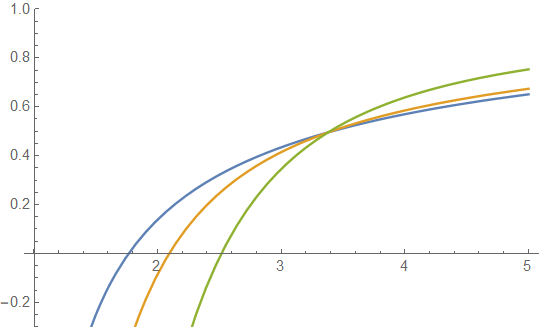}\qq
\includegraphics[width=7cm]{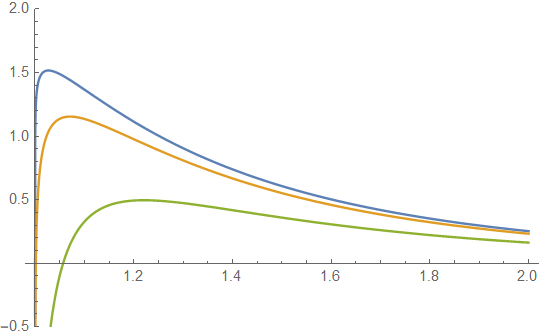}
\caption{\small 
	 The behavior of $A(r)$ (left) and $L(\cF(r))$ (right) in Example 2 with $m=1$ and $N = 0.1, 0.3, 1$ 
	 (upside down at small $r$) }
\label{bb-f-ex2}
\end{figure*}
% ------------------------------------------------------       

% ===========================================
\section{Conclusion}
% ===========================================   
   
   In this article that summarizes our talk presented at RUSGRAV-18, many of the results of
   this study are only declared, they will be described in more detail in our forthcoming 
   papers \cite{bb-we25}. This concerns, above all, the stability study of magnetic mirror stars 
   and a detailed derivation of solutions with NED sources. 
   
   Our results show some points of disagreement with \cite{bb-tops2, bb-tops3}, where (in our 
   terminology) the properties of magnetic mirror stars are studied, these points are 
   outlined at the end of Section 3 and deserve a further investigation.
   
   Considering NED as a possible source for magnetic mirror stars, we have formulated 
   a methodology of finding such solutions starting from assumptions on the form of
   $r(x)$ or $\Gamma(x)$. We also made a general observation that provided both these 
   functions are regular and positive, there {\it always\/} exist mirror star solutions that
   can be obtained by properly choosing the integration constant $N$ in \eqn{bb-A1}. 
   
\Funding{This work was supported by the RUDN University Project No. FSSF-2023-0003.}

\ConflictThey   
   
} %% end au-defs   
   
%\newpage
% ================================================

\end{document}